\documentclass[aps,twocolumn,showpacs,superscriptaddress,prl]{revtex4}
\usepackage{epsfig}
\usepackage{amssymb}
\usepackage{color}

\def\avg#1{\langle#1\rangle}

\def\be{\begin{equation}}       \def\ee{\end{equation}}
\def\bea{\begin{eqnarray}}      \def\eea{\end{eqnarray}}

\def\pp{\parallel}
 
\begin{document}
\title{Four-coloring model and frustrated superfluidity in the diamond lattice}
\author{Gia-Wei Chern}
\affiliation{Theoretical Division, Los Alamos National Laboratory, 
Los Alamos, NM 87545, USA}
\author{Congjun Wu}
\affiliation{Department of Physics, University of California, San Diego, 
CA 92093, USA}

\begin{abstract}
We propose a novel four-coloring model which describes 
``frustrated superfluidity'' of $p$-band bosons in the diamond optical lattice.
The superfluid phases of the condensate wavefunctions
on the diamond-lattice bonds are mapped to four distinct colors at low temperatures. 
The fact that a macroscopic number of states satisfy the constraints that four differently colored bonds meet at the same site
leads to an extensive degeneracy in the superfluid ground state at the classical level.
We demonstrate that the phase of the superfluid wavefunction as well as the orbital angular momentum 
correlations exhibit a power-law decay in the degenerate manifold that is 
described by an emergent magnetostatic theory with three independent 
flux fields. Our results thus provide a novel example of critical superfluid phase with algebraic order in three dimensions.
We further show that quantum fluctuations favor a N\'eel ordering of orbital angular
moments with broken sublattice symmetry through the 
order-by-disorder mechanism.
\end{abstract}
\pacs{05.50.+q, 67.85.Hj, 03.75.Lm}
\maketitle

Strongly frustrated systems are hosts to various complex orders, unusual
phases and elementary excitations. A well-studied example is the emergence
of critical Coulomb phases in systems ranging from water ice~\cite{ice}, 
pyrochlore magnets (spin ice)~\cite{spin-ice,quantum-ice} to $p$-band fermions in the optical
diamond lattice (orbital ice)~\cite{orbital-ice}. A common feature shared by these systems is the appearance of 
an extensively degenerate manifold due to a large number of unconstrained 
degrees of freedom. Although the macroscopic degeneracy is usually of geometrical 
origin, especially for magnetic systems~\cite{moessner2006}, 
the nontrivial spatial dependence of anisotropic orbital interactions in
optical lattices provides a new playground for studying interesting phenomena
related to a highly degenerate manifold~\cite{orbital-ice,wu2008,zhao2008,hauke2011}. 
In particular, for $p$-orbital bosonic condensates in optical lattices, the frustrated 
couplings between the $U(1)$ phase degrees of freedom 
can lead to intriguing quantum many-body states~\cite{wu2009,moore04}.

For conventional superfluid states of bosons,
their ground state wavefunctions are positive-definite
as stated by the ``no-node'' theorem which is valid under very general 
conditions \cite{feynman1972}.
It states that the superfluid phases are uniform over the entire condensate,
in other words, non-frustrated.
Exotic states of bosons beyond the ``no-node'' theorem have been theoretically 
proposed and experimentally realized in ultra-cold atom optical lattices~\cite{kuklov06,olschlager2011,choi11,li12}.
For example, when bosons are pumped to high orbital bands, their condensate 
wavefunctions form non-trivial 
representations of the lattice symmetry group, dubbed ``unconventional''
Bose-Einstein condensations, say, the complex-valued
$p+ip$-type condensates with spontaneously broken time-reversal symmetry
(see Ref.~\cite{wu2009} for a brief review).
These condensates are meta-stable excited states, and hence are not
constrained by the ``no-node'' theorem.

In this article we study the Bose-Einstein condensates formed by bosons pumped into the
$p$-orbital bands in a diamond optical lattice. We show that the problem of 
inter-site phase coherence in this lattice represents a highly frustrated system
and can be mapped to a four-coloring model on the diamond lattice.
By combining analytical arguments and numerical simulations,
we demonstrate the existence of an algebraic Coulomb phase in the highly degenerate
superfluid ground state at the classical level. 
Our work thus provides a three-dimensional (3D) generalization of the celebrated 
three-coloring model on the planar honeycomb lattice~\cite{baxter1970}.
Moreover, we also demonstrate a 3D superfluid phase with critical power-law
correlations.
We also show that quantum order-by-disorder selects a ground state with an 
quadrupled unit cell and a N\'eel ordering of the orbital angular moments.

{\em Model Hamiltonian.}
We begin with a discussion of the Hamiltonian for $p$-orbital bosons
in the diamond optical lattice. 
This bipartite lattice can be generated by the interference of four 
laser beams with suitably arranged light polarizations as discussed in 
Ref.~[\onlinecite{toader2004}]. 
The unit vectors from each site in sublattice $A$ to its four neighboring 
$B$-sites are denoted as $\hat\mathbf n_0=[111]$, $\hat\mathbf n_1=[1\bar 1\bar 1]$, 
$\hat\mathbf n_2=[\bar 1 1 \bar 1 ]$, and $\hat\mathbf n_3=[\bar 1 \bar 1 1]$.
Around the center of each lattice site, the point group symmetry is $T_d$ of
which the degenerate $p_{x,y,z}$-orbitals form a triplet irreducible representation.
The band energy is thus represented as
\begin{equation}
H_0= t_\pp\,\sum_{\nu=0}^3 \sum_{\langle ij \rangle \parallel \hat\mathbf n_\nu}
\left\{ p^\dagger_{i,\nu}\, p^{\phantom{\dagger}}_{j,\nu} 
+ \mbox{h.c.}\right\},
\label{eq:ham_kin}
\end{equation}
where $p_{\nu}$ is the projection of the $p$-orbitals
along the direction of $\hat\mathbf n_\nu$: $p_{\nu} = \hat\mathbf n_\nu\cdot\vec p$,
where $\vec p = (p_x, p_y, p_z)$ is a vector of annihilation operators for $p$-orbitals.
We only keep the longitudinal hopping between $p$-orbitals along the
bond direction, {\it i.e.}, the $\sigma$-bonding, and neglect the
small transverse hopping terms. Here
$t_\pp$ is positive because of the odd parity of $p$-orbitals~\cite{liu2006}.

The general on-site interaction has the form: 
$H_{\rm int} = \sum_i V_{ab;cd}\,\, p^\dagger_{i,a}\, p^\dagger_{i,b}\, p^{\phantom{\dagger}}_{i,c}\, 
p^{\phantom{\dagger}}_{i,d}$, where 
$V_{ab;cd}=g\int d^3 \mathbf r\, \psi_a (\mathbf r) \psi_b (\mathbf r ) 
\psi_c (\mathbf r) \psi_d (\mathbf r)$, $g$ is the contact interaction parameter,
and $\psi_a$ $(a=x,y,z)$ are Wannier functions of $p$-orbitals. 
The $T_d$ group contains the symmetry of rotation around 
the $z$-axis followed by the inversion, which transforms 
$\psi_z\rightarrow -\psi_z, \psi_{x}\rightarrow -\psi_y$, and
$\psi_y \rightarrow \psi_x$. Since $V_{ab;cd}$ should be invariant
under this transformation, symmetry consideration indicates that
there are only two independent parameters $V_{xx;xx}$ and $V_{xy;xy}$.
The interaction terms can thus be reorganized as 
\begin{equation}
H_{\rm int} = \frac{U}{2}\sum_{i} 
\Big\{ n_{i}^2-\frac{1}{3} \mathbf L_{i}^2 \Big\}
+\Delta U \sum_{i}\sum_{\alpha=x,y,z} n_{i, \alpha}^2, 
\label{eq:ham_int}
\end{equation}
where $n_{i,\alpha}$ is the particle number in $p_{\alpha}$-orbital ($\alpha = x, y, z$)
at site $\mathbf r_i$, $n_{i}=n_{i, x}+n_{i, y}+n_{i, z}$, and
$L_i^\mu=-\epsilon_{\mu\nu\lambda} p^\dagger_{\nu}p_{\lambda}$ 
is the orbital angular momemtum operator.
The first term in Eq.~(\ref{eq:ham_int}) represents the Hubbard 
interaction with the spherical symmetry around the site center
and $U=3V_{xy,xy}$ as given in Ref.~[\onlinecite{liu2006,wu2009}] .
We assume repulsive interactions $U>0$.
The ferro-orbital interaction means that bosons prefer to maximize
the on-site orbital angular momentum such that their wavefunctions
are most extended spatially to reduce the onsite repulsion~\cite{wu2009}, 
analogous to the second Hund's rule of
electrons filling in degenerate atomic orbitals. 
The second term with $\Delta U= \frac{1}{2} (V_{xx;xx}-3 V_{xy,xy})$ 
comes from the $T_d$ symmetry crystal field.
Compared with the spherically symmetric case ($\Delta U = 0$), the angular distributions
of Wannier orbitals $\psi_a$ in the diamond lattice expand from
$\pm \hat\mathbf  x$,$\pm \hat\mathbf y$,$\pm \hat\mathbf z$  toward the the bond directions
$\hat \mathbf n_\nu$.
This enhances the inter-orbital repulsion but weakens the intra-orbital
one, such that $\Delta U <0$.

\begin{figure}[t]
\includegraphics[width=0.92\columnwidth]{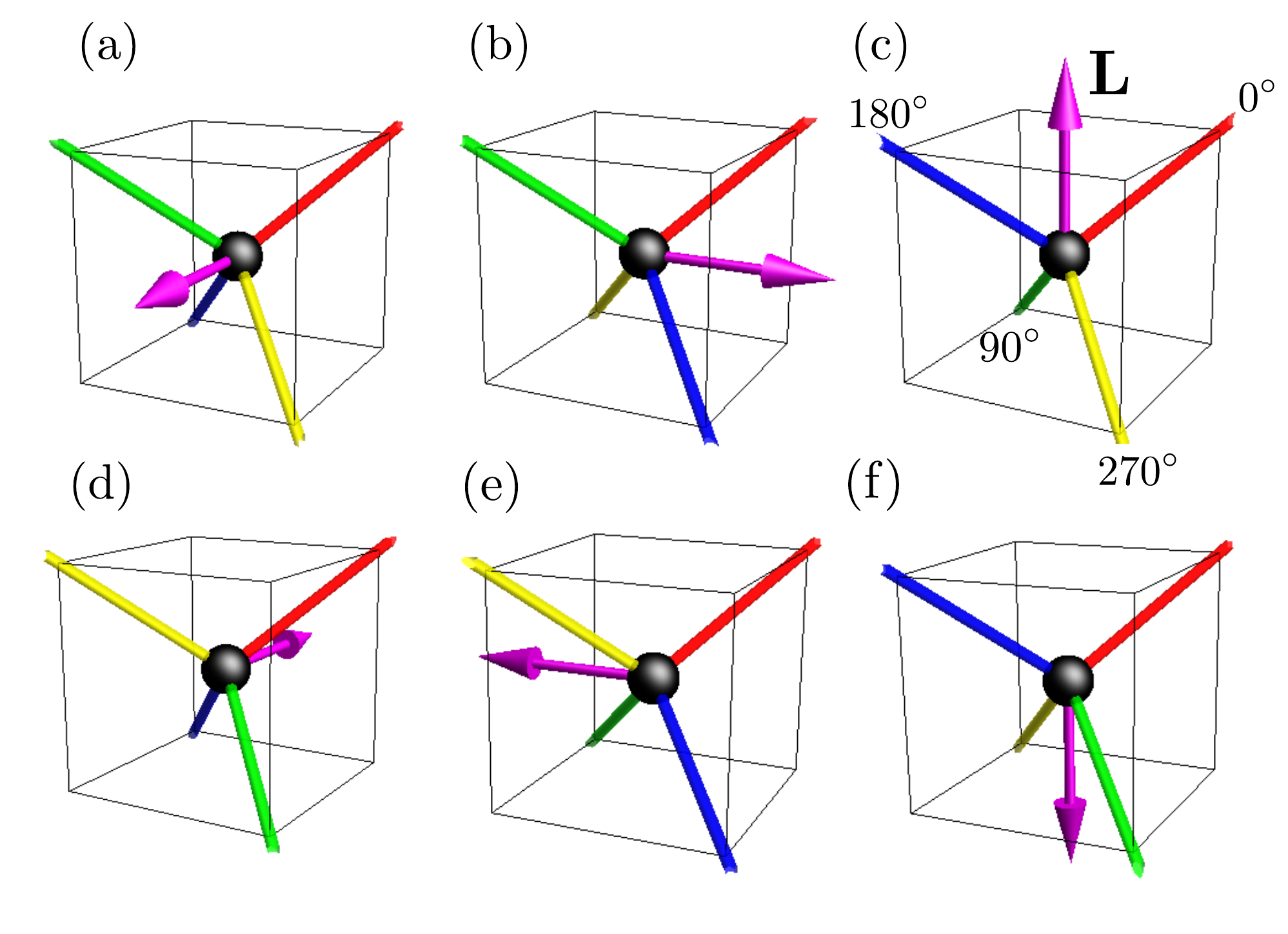}
\caption{(Color online) Mapping between the orbital angular momentum
and color configuration on the diamond lattice. The red, green, blue, 
and yellow bonds have phases $\phi = 0^\circ$, $90^\circ$, $180^\circ$,
and $270^\circ$, respectively, when projected to the base plane perpendicular
the direction of angular momentum $\mathbf L$. The mapping is not one-to-one,
as each $\mathbf L$ corresponds to 4 different cyclic 
permutations of the colors which preserve the same chirality.
\label{fig:mapping}}
\end{figure}

{\em Mapping to the 4-coloring model.}
We consider the strong coupling case with a weak crystal field $U\gg |\Delta U|$.
Assuming that the average filling number per site is large, we approximate 
each site as a small condensate and treat it classically. 
The $\Delta U<0$ term on each site favors a $p+ip$-type on-site condensate, 
which aligns $\mathbf L$ along
the cubic directions $\pm\hat x$, $\pm \hat y$, and $\pm \hat z$. 
In each site the direction of $\mathbf L$ is the nodal line of the 
condensate wavefunction, around which the $U(1)$ phase winds $2\pi$.

Let us consider site $i$ belonging to sublattice $A$. 
The projections of its four bond directions in the $xy$, $yz$ and $zx$ 
planes are evenly distributed.
Therefore, no matter which cubic direction $\mathbf L_i$ takes, 
the superfluid phase from site $i$ to its neighbor $j$ along the direction 
of $\hat\mathbf n_\nu$ can only be four different values among 
$0^\circ, 90^\circ, 180^\circ$, and $270^\circ$ up to an overall phase,
which is denoted as
\bea
\phi_{i,\nu}= \theta_i +\sigma_{i,\nu}\frac{\pi}{2},
\label{eq:phase}
\eea
where $\theta_i$ is the overall phase;
$\sigma_{i,\nu}=0,1,2$, and $3$ corresponding to colors {\bf R}, {\bf G}, 
{\bf B}, and {\bf Y}, respectively, in Fig.~\ref{fig:mapping}.
The phase shift $\theta_i \rightarrow \theta_i+\frac{\pi}{2}$
is equivalent to a cyclic permutation among the 4 colors around the 
direction of $\mathbf L_i$, thus $\theta_i$ on each site only takes values 
from $-\frac{\pi}{4}$ to $\frac{\pi}{4}$.
By the same reasoning, the phase from site $j$ on $B$ sublattice to 
its neighboring site $i$ along the
direction of $-\hat \mathbf n_\nu$ is denoted
as $\phi_{j,\bar \nu}= \theta_j +\sigma_{j,\bar\nu}\frac{\pi}{2}$.
The expectation value of the orbital operators appearing in the hopping 
Hamiltonian~(\ref{eq:ham_kin}) is given by 
$\langle p_{i, \nu} \rangle =  + \sqrt{n/3}\, e^{i\phi_{i,\nu}}$, 
and $\langle p_{j, \nu} \rangle =  - \sqrt{n/3}\, e^{i\phi_{j,\bar\nu}}$, for the two respective sublattices.
where $n$ is the average filling per site and the quantum depleting 
of the condensation fraction is neglected.
Thus the total kinetic energy summing up all the bonds reads
\bea
E 
&=&  -\frac{2}{3} n\, t_\pp \sum_{\langle ij \rangle \parallel \hat\mathbf n_\nu}
\cos \left[\theta_i-\theta_j - \frac{\pi}{2}(\sigma_{i,\nu}
-\sigma_{j,\bar\nu}) \right].
\eea

\begin{figure}[t]
\includegraphics[width=0.9\columnwidth]{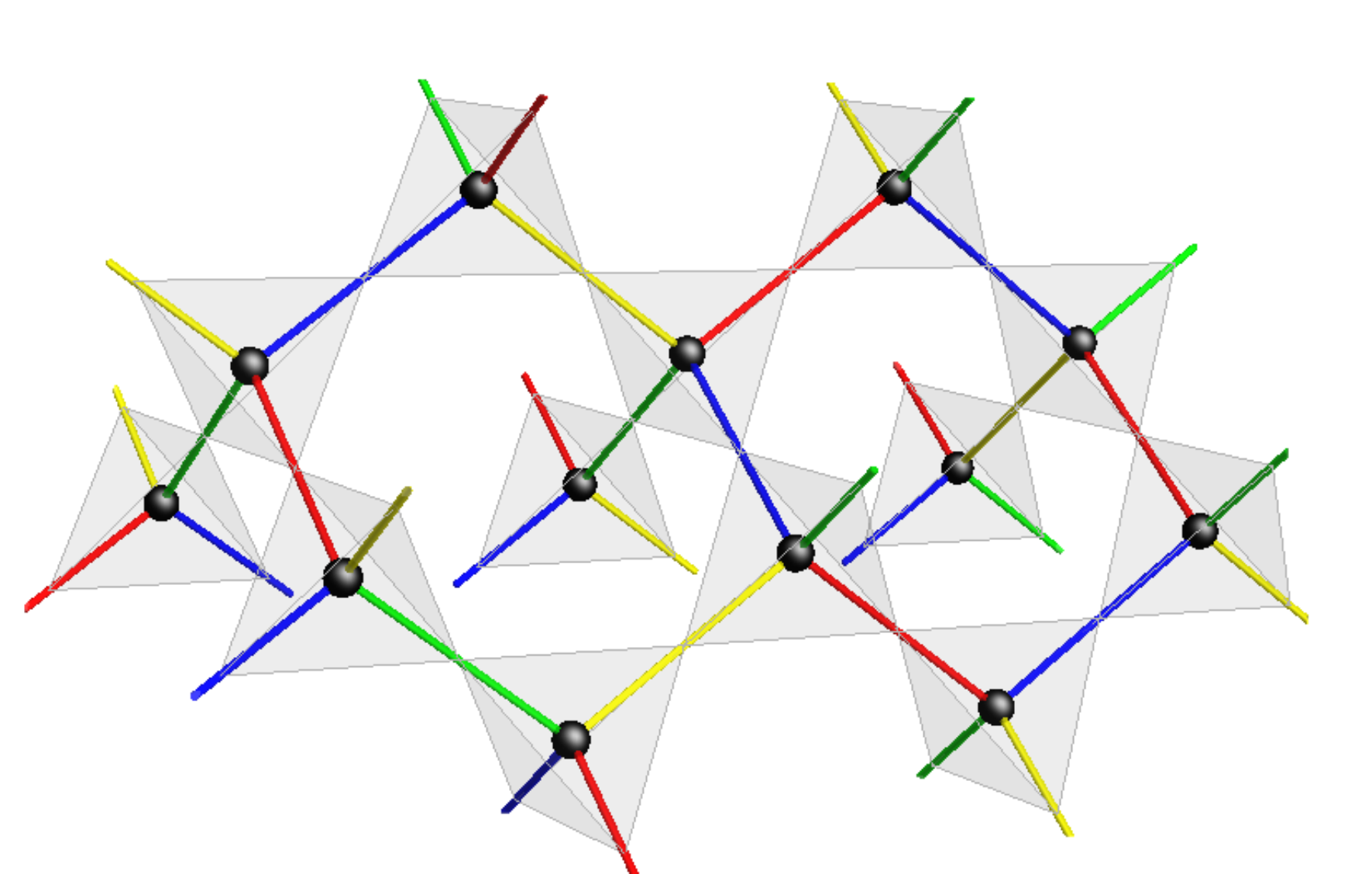}
\caption{(Color online) Coloring of the dimaond lattice
that satisfy the constraints that four bonds attached to the same vertex
have different colors. 
\label{fig:diamond}}
\end{figure}

To minimize the kinetic energy, we freeze $\theta_i = \theta_j =0$ for all sites,
and match the color on each bond, i.e. 
$\sigma_{i,\nu}=\sigma_{j,\bar \nu}=\sigma_{ij}$.
A defect occurs at a bond $\avg{ij}$ when $\sigma_{i,\nu}\neq \sigma_{j,\bar \nu}$.
The defect-free classic ground states of the $p$-band bosons thus map into 
a highly frustrated 4-coloring model~\cite{kondev1995}.  
An example of such configurations in the diamond lattice is shown 
in Fig.~\ref{fig:diamond}.
There are $4\,!=24$ coloring configurations on each site $i$, and $6$
different orientations $(\pm \hat\mathbf x, \pm \hat\mathbf y, 
\pm \hat\mathbf z)$ for $\mathbf L_i$, which are defined as the
``chirality'' of the condensate.
For each chirality, there are $24/6=4$ coloring configurations
corresponding to a $C_4$ rotation around the direction of $\mathbf L_i$; 
see Fig.~\ref{fig:mapping}. Consequently, the classical ground states
correspond to those of the 4-coloring model modular 4.
The chirality can be explicitly computed as $\mathbf L_i \parallel
(\hat \mathbf n_{\frac{\pi}{2}} - \hat \mathbf n_0)
\times (\hat \mathbf n_\pi - \hat \mathbf n_{\frac{\pi}{2}})$,
where $\hat \mathbf n_\varphi$ denotes the bond with phase $\varphi$ 
at site $\mathbf r_i$.

The phase correlation of the $p$-band condensate has contributions
from both the discrete color variables $\sigma_{ij}$ and the $U(1)$ phases $\theta_i$.
At the classic level, it is dominated by the color correlations at low 
temperatures in which the gapped excitations of defects can be neglected.
For all the defect-free configurations, the partition functions for 
thermal phase fluctuations are identical, and thus the color
correlations of $\sigma_{ij}$ and phase fluctuations of 
$\theta_i$ are decoupled.
In 3D, there exists a long-range-order 
for the $U(1)$ phase $\theta_i$ below a critical temperature $T_c$.
Consequently, up to a temperature-dependent factor, the correlations of phases $\phi_{i,\nu}$ 
are just described by the color correlations.

The residual entropy of the above 4-coloring model can be computed using a method similar to the 
Pauling estimation for the degeneracy of water ice \cite{ice}. Consider a given site on the diamond
lattice, there is a total of $4^4$ different colorings for the four bonds attached to it.
Only $4\,! = 24$ out of the $4^4$ coloring schemes satisfy the color constraint.
Treating the constraints imposed by different sites as independent, 
the number of degenerate ground states is $W \sim 4^N (4!/4^4)^{N/2}$, where $N$
is the number of nearest-neighbor bonds and $N/2$ is the number of sites on the diamond lattice.
This estimation gives an entropy density: $S_0/k_B = (1/2)\ln(3/2) \approx 0.2027$.
Since the 4-coloring model on diamond lattice can be mapped to an antiferromagnetic Potts model
on pyrochlore with infinite nearest-neighbor couplings~\cite{suppl}, the residual entropy can also be estimated
numerically with the aid of Monte Carlo simulations. The numerically obtained $S_0 \approx 0.2112\,k_B$
is very close to the Pauling estimation.


\begin{figure}[t]
\includegraphics[width=0.99\columnwidth]{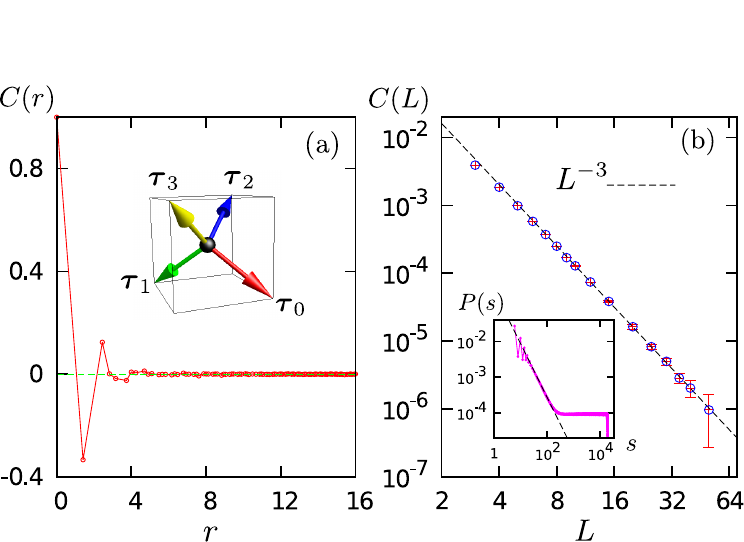}
\caption{(Color online) 
(a) The angular averaged superfluidy phase correlation function $C(r) = \langle e^{-i\phi(r)}\,e^{i\phi(0)}\rangle$ 
on a system with linear size $L=10$ in the ground state. 
The inset in (a) shows the four unit vectors ${\boldmath \mbox{$\tau$}}_{s}$ ($s = 0 \sim 3$) used in the definition
of the emergent magnetic fields.
(b) Phase correlation vs linear system size $L$. Here $C(L)$  denotes the color correlation between two points separated by $(L/2, L/2, 0)$ 
in a finite lattice containing $L^3$ cubic unit cells.
The inset shows the probability distribution $P(s)$ of flippable loop lengths in the critical phase obtained from simulations on a $L=50$ lattice.
\label{fig:simulation}}
\end{figure}

{\em Critical superfluid phase in 3D}.
We now investigate the phase correlations of the superfluid wavefunction in the degenerate manifold.
To this end, we employ Monte Carlo simulations with non-local loop updates to efficiently navigate the 4-coloring manifold~\cite{kondev1995,huse1992}.
In each loop update, two sites are chosen randomly in a given state of the manifold. 
These two sites will necessarily be of different color, say {\bf R} and {\bf B}. With periodic
boundary conditions, a {\bf RB}-colored loop containing the two chosen sites is
uniquely determined. By exchanging colors {\bf R} and {\bf B} over the length of the loop,
the non-local update results in a new state as all the color constraints
remain satisfied. 
Since all the microstates have equal statistical weight, loop updates of all lengths
and colors are accepted in order to satisfy detailed balance. 
The probability distribution of loop length $s$ shown in the inset of Fig.~\ref{fig:simulation}(b) exhibits
a power-law behavior $P(s) \sim s^{-3/2}$ for short loops, indicating a lack of length scales
in the degenerate manifold. The $s$-independent regime at large loop lengths is due to winding loops in a finite system
with periodic boundary conditions~\cite{jaubert2011}.

We then employ the above loop-update Monte Carlo simulations to investigate 
the correlations between the superfluid phases at 
different bonds: $C(r_{mn}) = \langle e^{-i\phi_m}\,e^{i\phi_n}\rangle$,
where $\phi_{m} = \phi_{i, \nu} = \phi_{j, \bar\nu}$, and the bond label $m = \langle ij \rangle$.
As discussed above, since the global $U(1)$ symmetry is broken below the critical temperature $T_c$,
the phase fluctuations are mainly governed by the correlations of the color variables $\sigma_m$
via Eq.~(\ref{eq:phase}). By averaging over $10^5$ allowed coloring configurations generated by the loop algorithm
in a $L=10$ lattice, we find a phase correlation $C(r)$, shown in Fig.~\ref{fig:simulation}(a), 
that falls off rapidly beyond a few nearest-neighbor distances,
indicating a color-disordered phase. Remarkably, the correlation function exhibits a power-law decay $C(L)\sim L^{-3}$ 
at long distances, as demonstrated in our large-scale Monte Carlo simulations with number of bonds up to $2\times 10^6$; see Fig.~\ref{fig:simulation}(b). This novel phase thus provides a rare example
of critical superfluid phase with algebraic order in three dimensions. Although this  phase is similar to the well known
 Kosterlitz-Thouless (KT) phase in 2D, the critical correlations in our case originate from the orbital frustration instead of thermal or quantum fluctuations
as in the KT scenario.

{\em Emergent Coulomb phase.}
The critical correlations of the $p$-band superfluid phases can be traced to the non-trivial local ordering imposed by the color constraints.
Similar to the case of dimer or spin-ice models~\cite{huse2003,isakov2004,henley2005,hermele2004},
these local constraints can be mapped to a conservation law of effective magnetic fluxes in the continuum approximation.
To this end, we first introduce four unit vectors ${\boldmath \mbox{$\tau$}}_{s}$ ($s = 0, 1, 2, 3$)
pointing toward different corners of a regular tetrahedron [inset of Fig.~\ref{fig:simulation}(a)] to represent the four different coloring. 
We can then construct three `magnetic' fields, each corresponds to a component of 
the ${\boldmath \mbox{$\tau$}}_{s} = (\tau^x_s, \tau^y_s, \tau^z_s)$ vectors, 
at the diamond sites:
\begin{eqnarray}
	\mathbf B^\alpha(\mathbf r_i) = \sum_{\nu=0}^3 \tau^\alpha_{\sigma_{i, \nu}}\,\hat\mathbf n_\nu,
\end{eqnarray}
where $\alpha = x, y, z$, and $\hat\mathbf n_\nu$ denotes the nearest-neighbor bond direction.
In the coarse-grained approximation, the color constraint that the four color variables $\sigma_{i, \nu}$
around a given site $i$ assume different values
translates to a divergence constraint $\nabla\cdot\mathbf B^{\alpha}(\mathbf r) = 0$ for the magnetic fields.

The effective free energy of the ground-state manifold arises entirely from entropy
and has the form of the magnetostatic theory with three independent flux fields
$
	\label{eq:F}
	F\bigl[\mathbf B^\alpha(\mathbf r)\bigr] \propto\sum_{\alpha}\int d^3\mathbf r
	 \left|\mathbf B^{\alpha}(\mathbf r)\right|^2.
$
Essentially, it states that the partition function
is dominated by microstates characterized by $\mathbf B^\alpha \approx 0$.
This is due to a large number of flippable loops in such states.
Although superficially $F$ describes Gaussian 
fluctuations of the magnetic fields, the divergence-free constraint in momentum 
space $\mathbf k\cdot\mathbf B^\alpha(\mathbf k)=0$ indicates that only
transverse fluctuations are allowed. Consequently, the correct correlators 
are obtained by projecting out the longitudinal fluctuations.
The asymptotic field correlators in real-space has the famous dipolar form
\begin{eqnarray}
	\langle B^\alpha_a(\mathbf r) B^\beta_b(0)\rangle \propto \delta_{\alpha\beta}
	(\delta_{ab} - 3\hat r_a \hat r_b)/r^3
\end{eqnarray} 
characteristic of a  Coulomb phase~\cite{henley2010}, which is confirmed in our
Monte Carlo simulations; see Fig.~\ref{fig:simulation}(d). 
In terms of the flux fields, the orbital angular momentum can be
expressed as $\mathbf L \sim \mathbf B^\alpha \times \mathbf B^\beta$,
where the two components $\alpha \neq \beta$ depend on the particular mapping between
phases and colors. The correlation function of angular momentum thus exhibits
a power-law $1/r^6$ decay in the degenerate manifold.
 
{\em Four-boson condensate superfluid order}.
While the critical phase described above does not have a single-boson long-range superfluid order,
there is a four-boson condensate order in this phase as the relative phase differences among {\bf R}, {\bf G}, {\bf B}, and {\bf Y} 
are integer multiples of $2\pi / 4$. 
Contrary to other reported multi-boson superfluid order~\cite{berg09,rad09,zhou01,mukerjee06,jian11} which arises from 
the interplay of attractive interactions and disordering fluctuations, the four-boson condensate in our case originates 
from the frustrated inter-site phase coherence, i.e. the kinetic energy, which is in turn induced by the anisotropic orbital hopping and geometrical frustration.

\begin{figure}[t]
\includegraphics[width=0.95\columnwidth]{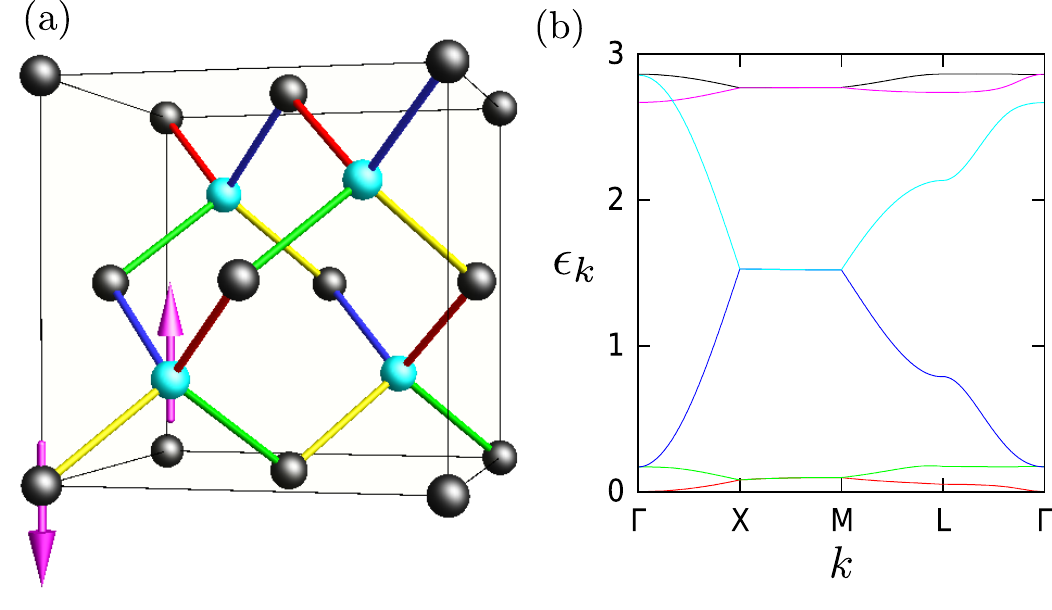}
\caption{(Color online) (a) N\'eel ordering of orbital moments $\mathbf L$ and
the corresponding 4-color configuration. The orbital moments $\mathbf L_i$ on the two sublattices
(labeled by dark and cyan balls) are antiparallel to each other.  (b) shows the band structure of
the Bogoliubov quasiparticles in the orbital N\'eel state ($U/t_{\parallel} = 0.1$).
\label{fig:neel-L}}
\end{figure}

{\em Quantum order-by-disorder}.
The extensive classical degeneracy of the 4-coloring manifold is removed by 
quantum fluctuations at the lowest temperatures. Here we consider the effects of 
zero-point motion energy of elementary excitations in different 4-color states.
To this end, we first restrict ourselves to long-range orders with a large
extended cubic unit cell containing 64 sites. This set of states include potential
orderings with high-symmetry wavevectors such as the $K$, $M$, or $L$ points.
We then consider fluctuations around the condensate wavefunction: $\vec p_i = \langle \vec p_i \rangle + \delta \vec p_i$.
The quasiparticle spectrum $\epsilon_{m}(\mathbf k)$ is then obtained by the standard Bogoliubov analysis~\cite{oosten01,cai12}.
By performing simulated annealing simulations to minimize the zero-point motion energy $\frac{1}{2}\sum_{m,\mathbf k} \epsilon_m(\mathbf k)$,
we find the state with a staggered arrangement
of the orbital moments $\mathbf L_i$ shown in Fig.~\ref{fig:neel-L}(a) is favored by quantum fluctuations.

The band structure of the Bogoliubov quasiparticles in this orbital N\'eel state is shown in Fig.~\ref{fig:neel-L}(b).
The explicit orbital wavefunctions on the two sublattices  
are given by $\langle \vec p_{A/B}(\mathbf r)\rangle = \frac{1}{2} \sqrt{n} (1, \pm i, 0)\,e^{i\mathbf Q\cdot\mathbf r}$ (up to an arbitrary $U(1)$ phase)
for the state shown in Fig.~\ref{fig:neel-L}(a).
Here the ordering wavevector $\mathbf Q = 2\pi (0, 0, 1)$ and $n$ is the particle density per site. 
The corresponding orbital moments $\mathbf L_{A/B} = \pm n \,\hat\mathbf z$ are uniform within the same sublattice
(a $\mathbf q = 0$ ordering), with the moments of the two sublattices antiparallel to each other.

The long-range ordering induced by quantum fluctuations occurs in both the diagonal and off-diagonal channels,
in other words, the (single-boson) superfluid phase ordering and the orbital angular momementum ordering coexist.
It is worth noting that since the four-boson off-diagonal order is not due to attractive interaction, there is no threshold (e.g. overcoming 
a color-disordering gap) for the quantum order-by-disorder mechanism to lift the degeneracy and select the ordered state shown in Fig.~\ref{fig:neel-L}(a).

{\em Conclusion}. 
We have studied a diamond-lattice four-coloring model which
describes the frustrated couplings between the superfluid phase
degrees of freedom for $p$-band Bose-Einstein condensates in the 
diamond lattice. We have also shown that the ground states of the 4-coloring
model is macroscopically degenerate and is described by an effective
magnetostatics theory with three independent flux fields. 
Both color and orbital angular momentum correlations decay
algebraically in this emergent Coulomb phase. Interestingly, point defects 
violating the color constraints carry ``magnetic'' charges associated with 
two of the three flux fields. A future direction of study is to 
explore the kinematics and dynamics of these novel quasiparticles. Finally,
we show that quantum order-by-disorder mechanism lifts the degeneracy
and favors N\'eel ordering of the orbital angular moments.

{\it Acknowledgment.} 
We thank useful discussions with R.~Moessner, Y. Li,  and Zi Cai.
G.W.C. acknowledges the support of ICAM and NSF Grant DMR-0844115.
C.W. is supported by the NSF DMR-1105945 and the 
AFOSR FA9550-11-1-0067(YIP).

{\it Note added.} Upon completion of this work, we became aware of
similar results on the pyrochlore 4-state Potts model in Ref.~\cite{khemani}.

\section{Supplementary Materials}

\begin{figure}
\includegraphics[width=0.8\columnwidth]{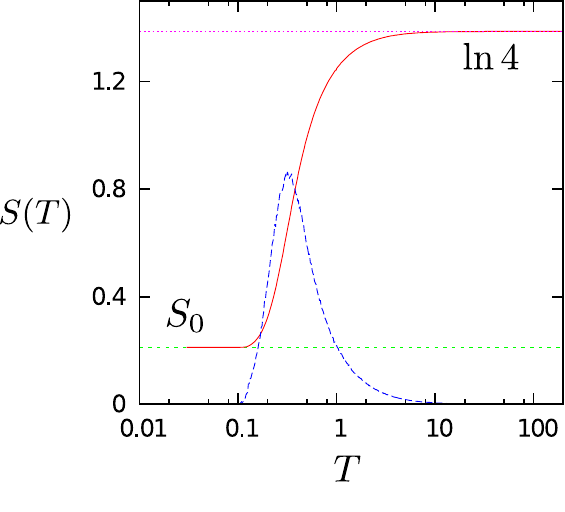}
\caption{(Color online) Monte Carlo simulations of the antiferromagnetic 4-state Potts model Eq.~(\ref{eq:H-potts})
on pyrochlore lattice: entropy $S$ and specific heat $C$ as a function of temperature $T$; 
the energy is measured in units of $J$. Fixing the high-$T$ entropy density to $k_B \ln 4$ gives a residual $S_0 \approx  0.2112 k_B$.
\label{fig:potts}}
\end{figure}

\subsection{Antiferromagnetic Potts model on pyrochlore lattice.}

Here we present a numerical estimate of the residual entropy of the 4-coloring model
on the diamond lattice.  Since the color variables are defined on bonds of the diamond lattice, 
it is more convenient to consider an equivalent Potts model on the pyrochlore lattice
whose sites correspond to the bond midpoints in the diamond lattice; see Fig.~1 in the main text. 
For convenience, we shall use $i$, $j, \cdots$ to denote the diamond sites, and use $m$, $n, \cdots$ 
for the sites on pyrochlore lattice.  A Potts variable $\sigma_m = \sigma_{ij} = 0, 1, 2, 3$ is assigned to each pyrochlore site such
that the corresponding phase is given by $\phi_m = \theta + \sigma_m \frac{\pi}{2}$, 
where $\theta$ is the global U(1) phase. 
The constraint that bonds attached to the same
vertex have different colors translates to an antiferromagnetic interaction
between nearest-neighbor Potts variables on the pyrochlore lattice:
\begin{eqnarray}
	H = J \sum_{\langle m l \rangle} \delta(\sigma_m, \sigma_l),
	\label{eq:H-potts}
\end{eqnarray}
Here the coupling constant $J \sim n\, t_{\parallel}$. To avoid confusion, we use $m, l$
to label the pyrochlore sites.
The hard constraint of the 4-coloring model is recovered in the $J \to \infty$ limit.
Consider first a single tetrahedron, there are $4\,!=24$ ground states in which
the four sites are in different Potts states. These correspond to the 24 coloring schemes 
on a diamond site. 

The pyrochlore Potts model~(\ref{eq:H-potts}) represents an under-constrained system with an extensively degenerate ground state, similar to their two-dimensional counterparts~\cite{baxter1970,kondev1995}.
The residual entropy of the 4-coloring model can be computed by performing finite
temperature Monte Carlo simulations of the pyrochlore Potts model~(\ref{eq:H-potts}). As shown in Fig.~\ref{fig:potts}(a),
by integrating the numerical specific-heat curve and fixing the high-$T$ entropy density to $k_B \ln 4$ (a Potts paramagnet),
we obtain a zero-temperature entropy density $S_0 \approx 0.2112\,k_B$.
The degeneracy can also be computed using a method similar to the 
Pauling estimation for the degeneracy of water ice \cite{ice}. Consider a single 
tetrahedron, $4\,!$ out of $4^4$ Potts configurations satisfy the color constraint.
Treating the constraints imposed by different tetrahedra as independent, 
the number of degenerate ground states is $W \sim 4^N (4!/4^4)^{N/2}$, where $N$
is the number of Potts variables and $N/2$ is the number of tetrahedra (or diamond sites).
This estimation gives an entropy density: $S_0/k_B = (1/2)\ln(3/2) \approx 0.2027$,
which is very close to the numerical value $S_0 \approx 0.2112\,k_B$.

\subsection{Mapping to emergent flux fields.}

To map the highly constrained 4-coloring manifold to the emergent flux fields, and to make manifest the analogy with the frustrated
spin systems, we consider a Heisenberg antiferromagnet on the pyrochlore lattice:
\begin{eqnarray}
	\label{eq:heisenberg}
	H = \mathcal{J} \sum_{\langle ml \rangle} \mathbf S_m\cdot\mathbf S_l 
	+ D \sum_m f(\mathbf S_m; \{{\boldmath \mbox{$\tau$}}_s\}).
\end{eqnarray}
Here $\mathcal{J}>0$ and $\mathbf S_m$ denotes a classical $O(3)$ vector of unit length.
The second term represents a special `tetrahedral' anisotropy:
the function $f(\mathbf S)$ has four degenerate minimum at directions ${\boldmath \mbox{$\tau$}}_s$
pointing toward different coners of a regular tetrahedron; see the inset of Fig.~3(a) in the main text.  
In the $D \to \infty$ limit, Hamiltonian~(\ref{eq:heisenberg}) reduces to the
Potts model as the spins are aligned to the tetrahedral vectors according 
to $\mathbf S_m = {\boldmath \mbox{$\tau$}}_{\sigma_m}$, where $\sigma_m$ is the corresponding Potts variable at site $m$.
The exchange term in~(\ref{eq:heisenberg}) can be rewritten as $(\mathcal{J}/2)\sum_{\boxtimes}\left|\mathbf S_{\boxtimes}\right|^2$
up to an irrelevant constant, where $\mathbf S_{\boxtimes}  = \sum_{m \in \boxtimes} \mathbf S_m$ denotes the total spin of a tetrahedron. 
The ground state of the above spin model is reached when the total spin $\mathbf S_{\boxtimes} = 0$ for all tetrahedra.

Since the Potts variables $\sigma_m$ in any individual tetrahedra assume four different values for a valid 4-coloring configuration on the diamond lattice,
and noting that the four  ${\boldmath \mbox{$\tau$}}_s$ vectors sum to zero, we have:
\begin{eqnarray}
	\mathbf S_{\boxtimes} = \sum_{m \in \boxtimes} \mathbf S_m = \sum_{m \in \boxtimes} {\boldmath \mbox{$\tau$}}_{\sigma_m} = 0,
\end{eqnarray}
for all tetrahedra, indicating the corresponding magnetic state $\{\mathbf S_m\}$ is a ground state of the spin model.
We thus establish an one-to-one correspondence between the 4-coloring configuration and the ground states of the spin model.

The mapping to the spin model~(\ref{eq:heisenberg}) also allows us to recast the
color constraint into a conservation law of effective magnetic flux similar to the case of pyrochlore spin model discussed in 
Refs.~\cite{isakov2004,henley2005,hermele2004}. 
We first define three `magnetic' fields, each corresponds to a component of the Heisenberg spin, on the diamond site
\begin{eqnarray}
	\mathbf B^\alpha(\mathbf r_i) = \sum_{\nu = 0}^3 S^\alpha_{\langle ij \rangle}\,\hat\mathbf n_\nu 
	= \sum_{\nu=0}^3 \tau^\alpha_{\sigma_m}\,\hat\mathbf n_\nu.
\end{eqnarray}
Here $\alpha = x, y, z$, the summation is over the four nearest-neighbor bond directions $\hat\mathbf n_\nu$, 
and the pyrochlore site $m$ corresponds to the bond $\langle ij \rangle \parallel \hat\mathbf n_\nu$ in the diamond lattice.
Since for every tetrahedra in the ground state, there are two $+$ and two $-$ signs for each component $\alpha$ 
of the four ${\boldmath \mbox{$\tau$}}_{\sigma_m}$ vectors. Thus the $\mathbf B$ field for {\em each} spin component $\alpha$
corresponds to a two-in-two-out configuration, i.e. the flux fields are conserved in each tetrahedra (no source or sink).
In the coarse-grained approximation, the color constraints $\mathbf S_{\boxtimes} = 0$ translate
to a divergence constraint $\nabla\cdot\mathbf B^{\alpha}(\mathbf r) = 0$ for the magnetic fields~\cite{henley2010}.

\end{document}